\documentclass[journal=inoraj,manuscript=article]{achemso}
\usepackage[version=3]{mhchem}
\usepackage{color}

\title{Giant thermoelectric power factor anisotropy in PtSb$_{1.4}$Sn$_{0.6}$}
\author{Yu Liu,$^{\dag,\sharp,\ast}$ Zhixiang Hu,$^{\dag,\ddag}$ Michael O. Ogunbunmi,$^{\S}$ Eli Stavitski,$^{\|}$ Klaus Attenkofer,$^{\|,\P}$ Svilen Bobev,$^{\S}$ and Cedomir Petrovic $^{\dag,\ddag,\ast}$}
\email{yuliu@lanl.gov; petrovic@bnl.gov}
\affiliation[BNL]
{$^{\dag}$Condensed Matter Physics and Materials Science Department, Brookhaven National Laboratory, Upton, New York 11973, USA\\
$^{\ddag}$Materials Science and Chemical Engineering Department, Stony Brook University, Stony Brook, New York 11790, USA\\
$^{\S}$Department of Chemistry and Biochemistry, University of Delaware, Newark, Delaware 19716, USA\\
$^{\|}$National Synchrotron Light Source II, Brookhaven National Laboratory, Upton, New York 11973, USA}

\begin{document}

\begin{abstract}
  We report giant anisotropy in thermoelectric power factor ($S^2\sigma$) of marcasite structure-type PtSb$_{1.4}$Sn$_{0.6}$. PtSb$_{1.4}$Sn$_{0.6}$, synthesized using ambient pressure flux growth method upon mixing Sb and Sn on the same atomic site, is a new phase different from both PtSb$_2$ and PtSn$_2$ that crystallize in the cubic $Pa\bar{3}$ pyrite and $Fm\bar{3}m$ fluorite unit cell symmetry, respectively. Giant difference in $S^2\sigma$ for heat flow applied along different principal directions of the orthorhombic unit cell mostly stems from anisotropic Seebeck coefficients.
\end{abstract}


\maketitle

\section{Introduction}

Anisotropy in functional characteristics is rather important in many diverse scientific disciplines including thermoelectricity.\cite{GallagherJ,HeY,RauziM,McManusI,KhajetooriansA,ObergJ,YanY,ZhaoL} Thermoelectric materials are of interest for power generation without greenhouse gas emissions as they convert heat into electric power and \emph{vice versa} and are characterized by the figure of merit $ZT$ at temperature $T$.\cite{Snyder} For high figure of merit $ZT= (S^{2}$$\sigma$/$\kappa$)$T$, where $S$ is thermopower, $\sigma$ and $\kappa$ are electrical and thermal conductivity, respectively: high thermoelectric power factor ($S^2\sigma$) and low thermal conductivity $\kappa$ are important. Whereas anisotropy in $S^2\sigma$ and $\kappa$ has been used in the past to boost thermoelectric performance, the observed anisotropy was modest, up to about an order of magnitude for heat flow along different principal axes of the crystal structure in known materials.\cite{PanY,GuilmeauE,KyratsiT,LiA}

Marcasites often feature high values of thermopower and thermoelectric power factor; in FeSb$_2$ thermoelectric properties can be further defect-tuned and are somewhat anisotropic.\cite{Bentien,Takahashi,Sales2,DuQ} Here we report giant thermoelectric power factor anisotropy in orthorhombic marcasite-type PtSb$_{1.4}$Sn$_{0.6}$. PtSb$_2$ and PtSn$_2$ are known to crystallize in a cubic structure of $Pa\bar{3}$ pyrite and $Fm\bar{3}m$ fluorite unit cell symmetry, respectively.\cite{Thomassen,Wallbaum} PtSb$_2$ features large thermopower and enhanced power factor due to the flat band in valence band dispersion.\cite{NishikuboY,MoriK,SaeedY} PtSn$_2$, on the other hand, is a metal with good catalytic performance.\cite{GuptaA,SuH} Even though crystallographic unit cell of end members are cubic, orthorhombic marcasite-type crystal structure was detected in the Pt-Sb-Sn ternary phase diagram.\cite{Bhan}

\section{EXPERIMENTAL}

\subsection{Synthesis}

A large PtSb$_{1.4}$Sn$_{0.6}$ single crystal with dimensions $\sim 8\times5\times3$ mm$^3$ was grown by the self-flux technique starting from an intimate mixture of pure elements Pt, Sb, Sn with a molar ratio of 0.15 : 0.425 : 0.425. Starting materials were sealed in an evacuated fused silica tube. The tube was heated to 1150 $^\circ$C over 20 h, kept at 1150 $^\circ$C for 3 h, then slowly cooled to 580 $^\circ$C at a rate of 1 $^\circ$C/h.

\subsection{Structure and Composition Analysis}

Powder x-ray diffraction (XRD) data were taken with Cu $K_{\alpha}$ ($\lambda = 1.5418$ {\AA}) radiation of a Rigaku Miniflex powder diffractometer. Element analysis was performed using energy-dispersive x-ray spectroscopy (EDS) in a JEOL LSM-6500 scanning electron microscope. The chemical composition determined by the SEM-EDS gives Pt : Sb : Sn = $32(1):46(2):22(2)$. Single crystal XRD data were collected on crystal mounted on a low background plastic holder using Paratone-N oil under the cold stream of nitrogen gas. Data acquisitions took place on a Bruker APEX-II CCD diffractometer using graphite monochromatized Mo $K_{\alpha}$ radiation ($\lambda$ = 0.71073 {\AA}). The data collection, data reduction and integration, as well as refinement of the unit cell parameters were carried out using the Bruker software. Semi-empirical absorption correction was applied with the aid of the SADABS software package. The structure was subsequently solved by direct methods and refined on $F^2$ with the aid of the SHELXL package.\cite{SHELXL} All atoms were refined with anisotropic displacement parameters with scattering factors (neutral atoms) and absorption coefficients.\cite{Tables} X-ray absorption spectroscopy (XAS) was measured at 8-ID beamline of the NSLS II at Brookhaven National Laboratory (BNL) in fluorescence mode. X-ray absorption near edge structure (XANES) and extended x-ray absorption fine structure (EXAFS) spectra were processed using the Athena software package. The EXAFS signal, $\chi(k)$, was weighed by $k^2$ to emphasize the high-energy oscillation and then Fourier-transformed in $k$ range from 2 to 10 {\AA}$^{-1}$ to analyze the data in $R$ space.

\subsection{Thermal Transport and Thermodynamic Measurements}

Transport, thermal and thermodynamic properties were measured in a quantum design PPMS-9 system. The crystal was oriented and cut into rectangular bars along principal crystallographic axes for four-probe resistivity and thermal transport measurements. Silver paint and thin Au wires were used for contacts in electrical transport measurement. The drive current and frequency were 1 mA and 17 Hz. Hall resistivity $\rho_{xy}$ was measured with current along the $c$-axis and magnetic field along the $b$-axis direction. Thermal transport was measured by using one-heater-two-thermometer setup. Epoxy and Cu leads were used for contacts in thermal transport measurement. Continuous measuring mode was used with maximum heat power and period set at 50 mW and 1430 s along with the maximum temperature rise of 3$\%$. The sample dimensions were measured by an optical microscope Nikon SMZ-800 with 10 $\mu$m resolution.

\section{RESULTS AND DISCUSSION}

\subsection{Structure analysis}

\begin{figure}
\centerline{\includegraphics[scale=1]{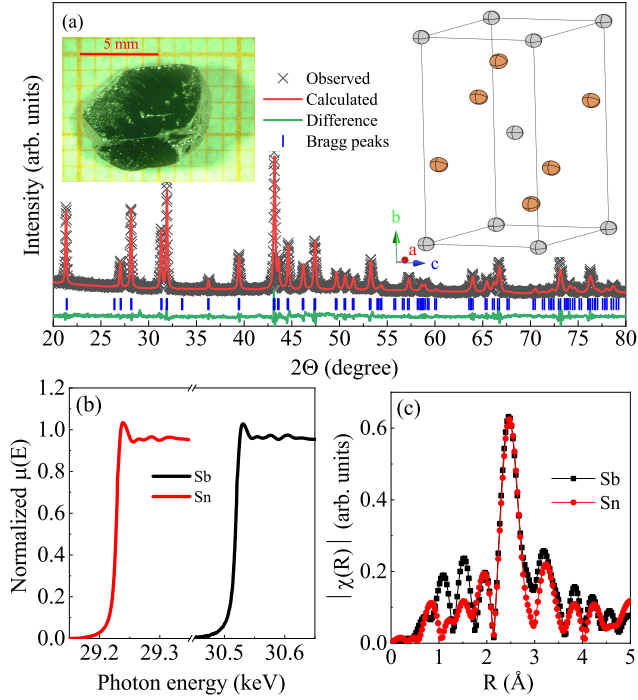}}
\caption{(Color online) (a) Powder x-ray diffraction spectra of pulverized PtSb$_{1.4}$Sn$_{0.6}$ single crystal. Experimental data are shown by ($\times$), structural model is represented by the red solid line, and difference curve (green solid line) is offset for clarity. The vertical tick marks represent Bragg reflections of the $Pnnm$ space group. Insets in (a) show the FeAs$_2$-type orthorhombic crystal structure with thermal elipsoids from single crystal x-ray refinement and a photo of one grown crystal on 1 mm grid. (b) Normalized Sn (left) and Sb (right) $K$-edge XANES spectra. (c) Fourier transform magnitudes of Sb and Sn $K$-edge EXAFS data measured at room temperature.}
\end{figure}

PtSb$_{1.4}$Sn$_{0.6}$ crystalizes in the FeAs$_2$-type structure, space group $Pnnm$ [Fig. 1(a)]. The determined lattice parameters from powder x-ray unit cell refinements are $a$ = 5.35(2) {\AA} $b$ = 6.60(2) {\AA}, and $c$ = 3.93(2) {\AA} at room temperature, in line with the previous report.\cite{Bhan} When compared to the single crystal x-ray measurement (Table I) at 200 K, the compression of $a$ and $b$ are $\sim$ 0.30\% and 0.33\%, respectively, three times larger than that of $\sim$ 0.10\% of $c$ axis lattice parameter.

\begin{table}\centering
\caption{\label{tab}Atomic coordinates, displacements parameters $U_{\textrm{eq}}$ (\AA$^2$) and occupancy (occ.) of atomic sites for PtSb$_{1.4}$Sn$_{0.6}$ refined from single crystal x-ray diffraction experiment taken at 200(2) K. $U_{\textrm{eq}}$ is defined as a trace of the orthogonalized $U_{\textrm{i,j}}$. The Sb/Sn ratio was fixed to the EDS values due to very small difference in atomic mass. The refined unit cell parameters are $a$ = 5.334(2) {\AA}, $b$ = 6.578(2) {\AA}, $c$ = 3.926(2) {\AA}.}
\begin{tabular}{ccccccc}
\hline\hline
Atom & Site & $x$ & $y$ & $z$ & $U_{\textrm{eq}}$ & occ. \\
\hline
Pt & 2$a$ & 0 & 0 & 0 & 0.00735(2) & 1 \\
Sb & 4$g$ & 0.20685(10) & 0.36427(7) & 0 & 0.0084(2) & 0.68 \\
Sn & 4$g$ & 0.20685(10) & 0.36427(7) & 0 & 0.0084(2) & 0.32 \\
\hline\hline
\end{tabular}
\end{table}

Figure 1(b) shows the normalized Sb and Sn $K$-edge XANES spectra. The local environment of Sb and Sn atoms is revealed in the EXAFS spectra measured at room temperature [Fig. 1(c)]. In a single-scattering approximation, the EXAFS can be described by: $\chi(k) = \sum_i\frac{N_iS_0^2}{kR_i^2}f_i(k,R_i)e^{-\frac{2R_i}{\lambda}}e^{-2k^2\sigma_i^2}sin[2kR_i+\delta_i(k)]$ \cite{Prins}, where $N_i$ is the number of neighbouring atoms at a distance $R_i$ from the photoabsorbing atom. $S_0^2$ is the passive electrons reduction factor, $f_i(k, R_i)$ is the backscattering amplitude, $\lambda$ is the photoelectron mean free path, $\delta_i$ is the phase shift, and $\sigma_i^2$ is the correlated Debye-Waller factor measuring the mean square relative displacement of the photoabsorber-backscatter pairs. The main peak at $R \sim 2.5$ {\AA} in the Fourier transform magnitudes of Sb and Sn $K$-edge EXAFS corresponds to the Pt-Sb/Sn bond distance $\sim$ 2.64(2.66) {\AA}. Basic units of the crystal structure are Pt(Sb/Sn)$_6$ octahedra where two apical distances are shorter than planar, similar to FeSb$_2$.\cite{Zaliznyak}

\subsection{Electrical Resistivity}

\begin{figure}
\centerline{\includegraphics[scale=1]{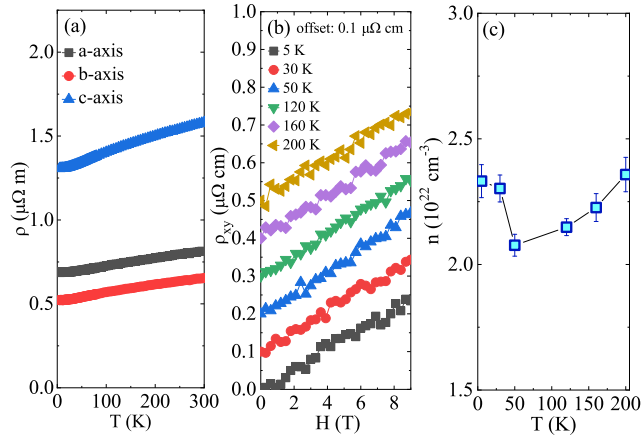}}
\caption{(Color online) (a) Temperature dependence of electrical resistivity $\rho$(T) along all three principal crystallographic axes for PtSb$_{1.4}$Sn$_{0.6}$ single crystal. (b) Field dependence of Hall resistivity $\rho_{xy}$(H) and (c) the derived carrier concentration $n$.}
\end{figure}

Figure 2(a) shows the temperature dependence of electrical resistivity $\rho$(T) along all three axes for PtSb$_{1.4}$Sn$_{0.6}$ crystal. A bad metal behavior is observed with a relatively low residual resistivity ratio (RRR = $\rho_{\textrm{300 K}}/\rho_{\textrm{4 K}}$) $\sim$ 1.18 - 1.25, and considerable anisotropy when the current flow is applied to different axis of the crystal structure, up to three times difference. The Hall resistivity $\rho_{\textrm{xy}}$ is linear in magnetic field [Fig. 2(b)], indicative of a dominant single-band transport; temperature-dependent Hall coefficient $R_\textrm{H}$ = $1/ne$ can be estimated from the linear fit and the derived carrier concentration $n$ is plotted in Fig. 2(c). The $n \sim 2.07(4) - 2.36(7) \times 10^{22}$ cm$^{-3}$ shows weak temperature dependence with a kink around 50 K.

\subsection{Thermal Conductivity}

\begin{figure}
\centerline{\includegraphics[scale=1]{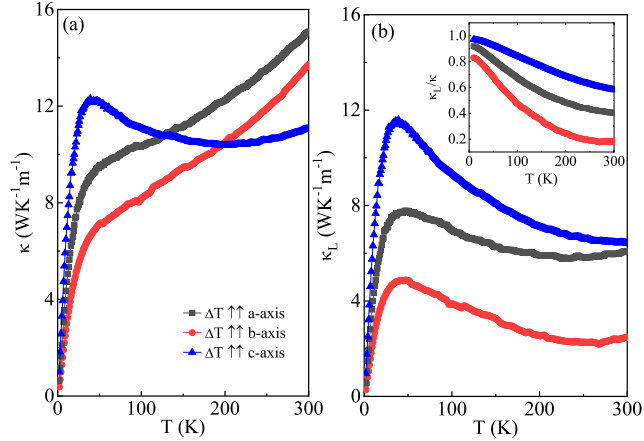}}
\caption{(Color online) Temperature dependence of (a) total thermal conductivity $\kappa$(T) and (b) phonon term $\kappa_\textrm{L}$(T) for thermal gradient $\Delta$$T$ applied along all three principal axes of the orthorhombic marcasite structure for PtSb$_{1.4}$Sn$_{0.6}$. Inset in (b) shows the ratio of lattice part $\kappa$$_\textrm{L}$/$\kappa$(T).}
\end{figure}

Figure 3(a) shows the temperature dependence of total thermal conductivity $\kappa$(T) for the heat flow applied along all three axes of the crystal structure. In general, $\kappa = \kappa_\textrm{e} + \kappa_\textrm{L}$, i.e. it includes contributions of the electronic part $\kappa_\textrm{e}$ and the phonon term $\kappa_\textrm{L}$. The $\kappa_\textrm{L}$ can be obtained by subtracting the $\kappa_\textrm{e}$ part calculated from the Wiedemann-Franz law $\kappa_\textrm{e}/T = L_0/\rho$, where $L_0$ = 2.45 $\times$ 10$^{-8}$ W $\Omega$ K$^{-2}$ and $\rho$ is the measured resistivity. The estimated $\kappa_\textrm{L}$ are depicted in Fig. 3(b). The $\kappa_\textrm{L}$ gradually dominates as temperature decreases, as shown by the ratio $\kappa_\textrm{L}/\kappa$ [Fig. 3(b) inset]. The anisotropic thermal conductivity in PtSb$_{1.4}$Sn$_{0.6}$ provides a good platform to theoretically investigate the anisotropic model for the minimum thermal conductivity ($\kappa_{min}$) \cite{kmin}, besides the Cahill-Pohl/Einstein model of $\kappa_{min}$ assuming isotropic properties \cite{DS}.

\subsection{Specific Heat and Thermopower}

\begin{figure}
\centerline{\includegraphics[scale=1]{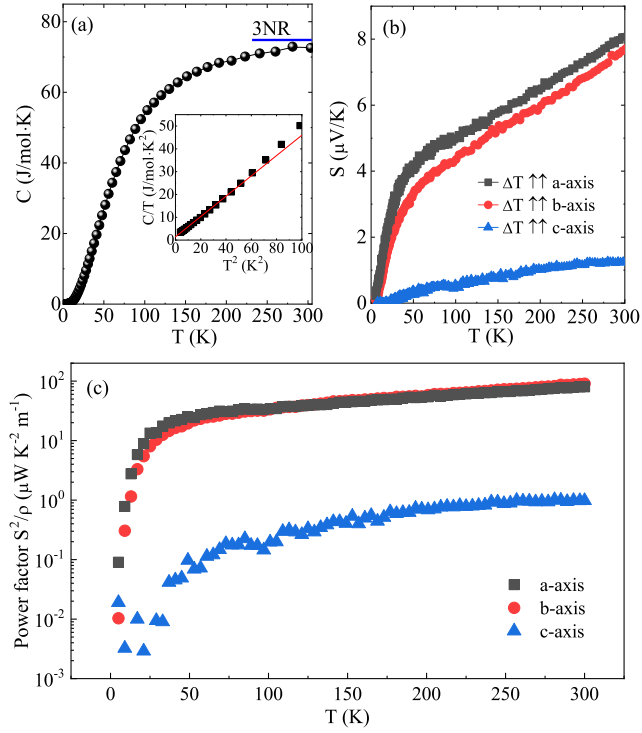}}
\caption{(Color online) (a) Temperature dependence of heat capacity $C_\textrm{p}$(T) for PtSb$_{1.4}$Sn$_{0.6}$ single crystal. Inset shows the low temperature data $C_\textrm{p}/T$ as a function of $T^2$ with the fitting by using $C_\textrm{p}/T=\gamma+\beta T^2$. (b) Temperature dependence of thermopower $S$(T) measured for thermal gradient applied along all three principal axes of the orthorhombic structure. (c) Temperature dependence of anisotropy in thermoelectric power factor $S^2$/$\rho$.}
\end{figure}

The specific heat $C_\textrm{p}$(T) of PtSb$_{1.4}$Sn$_{0.6}$ single crystal [Fig. 4(a)] approaches the value of $3NR$ at room temperature, where $N = 3$ is the atomic number in the chemical formula and $R$ = 8.314 J mol$^{-1}$ K$^{-1}$ is the gas constant, consistent with the Dulong-Petit law. The low temperature $C_\textrm{p}$(T) can be well fitted by using $C_\textrm{p}/T=\gamma+\beta T^2$ [inset in Fig. 4(a)]. The derived electronic specific heat coefficient $\gamma$ $\sim$ 1.41(9) mJ/mol-K$^2$ and the Debye temperature $\Theta_\textrm{D}$ $\sim$ 236(2) K from $\beta$ $\sim$ 0.446(4) mJ/mol-K$^4$ using the relation $\Theta_\textrm{D} = (12\pi^4NR/5\beta)^{1/3}$.

Figure 4(b) shows the temperature dependence of thermopower $S$(T) for the heat flow applied along all three principal axes of the marcasite structure. The sign of $S$(T) in the whole temperature range is positive, indicating hole-type carriers. This is consistent with the Hall resistivity analysis. The $S$(T) is almost linear at high temperature and changes its slope below 50 K. At low temperatures, the diffusive Seebeck response of Fermi liquid dominates and is expected to be linear in temperature.

In a single-band system, $S$(T) is given by $\frac{S}{T} = \pm\frac{\pi^2}{3}\frac{k_\textrm{B}^2}{e}\frac{N(\varepsilon_\textrm{F})}{n}$,\cite{Barnard,Miyake} where $k_\textrm{B}$ is the Boltzmann constant, $N(\varepsilon_\textrm{F})$ is the density of states at the Fermi energy $\varepsilon_\textrm{F}$, and $n$ is the carrier concentration. The electronic specific heat can be expressed as $\gamma = \frac{\pi^2}{3}k_\textrm{B}^2N(\varepsilon_\textrm{F})$. Therefore $S/T = \pm \gamma/ne$, where the units are V K$^{-1}$ for $S$, J K$^{-2}$ m$^{-3}$ for $\gamma$, and m$^{-3}$ for $n$, respectively. In general, we express $\gamma $ in J mol$^{-1}$ K$^{-2}$. Then we can define a dimensionless quantity $q=\frac{S}{T}\frac{N_\textrm{A}e}{\gamma}$, where $N_\textrm{A}$ is the Avogadro number, gives the number of carriers
per formula unit (proportional to 1/n). The derived $S/T$ from 5 to 30 K is $\sim$ 0.103(1) and 0.128(2) $\mu$V K$^{-2}$ for $a$- and $b$-axis, respectively, and 0.007(1) $\mu$V K$^{-2}$ for $c$-axis; the calculated $q$ is $\sim$ 7.0(4) and 8.7(4) for $a$- and $b$-axis, respectively, and 0.48(4) for $c$-axis of the orthorhombic marcasite structure for PtSb$_{1.4}$Sn$_{0.6}$. Given the volume of unit cell $\sim$ 0.139 nm$^3$, we calculate the carrier density per volume $n \approx 1.5(1) \times 10^{22}$ cm$^{-3}$ for the $c$-axis, close to the value obtained from the Hall measurement.

\subsection{Thermoelectric Power Factor}

Thermoelectric power factor $S^2\sigma$ where $\sigma$ = 1/$\rho$ is plotted in Fig. 4(c). The $S^2\sigma$ maximum below 300 K for the heat flow applied along $a$- or $b$-axis is smaller to what was observed in weakly defective FeSb$_2$ crystals without metal-insulator transition.\cite{Bentien,DuQ,JieQ} On the other hand, the maximum appears at room temperature with positive slope. This is far above the temperature of colossal $S^2\sigma$ value for FeSb$_2$ which is observed at 20 - 30 K, and suggests possible higher thermoelectric power factor above room temperature. Furthermore, $S^2\sigma$ is strongly anisotropic: heat flow applied along the $c$ axis results in about two orders of magnitude difference when compared to $a$ or $b$ direction [Fig. 4(c)]. The calculated values of $ZT$ at room temperature are $\sim$ 1.6$\times$10$^{-3}$ and 2.0$\times$10$^{-3}$ for $a$ and $b$ axis, respectively, which are also about two orders of magnitude larger than $\sim$ 2.6$\times$10$^{-5}$ for the $c$ axis of PtSb$_{1.4}$Sn$_{0.6}$. The absolute values are relatively small for thermoelectric application. The $ZT$ of nominal PtSb$_2$ reaches a maximum $\sim$ 0.009 between 220 to 250 K, which can be enhanced to $\sim$ 0.059 - 0.073 around 350 K in samples with 2\% Te-doping at Sb sites \cite{PtSb2}.

\section{CONCLUSION}

In summary, PtSb$_{1.4}$Sn$_{0.6}$ single crystals crystallize in orthorhombic marcasite structure even though both PtSb$_2$ and PtSn$_2$ are cubic materials at ambient pressure. The alloying in between PtSb$_2$ and PtSn$_2$ creates not only simple random distribution of atoms but also significant change in bond energies and ethalpies of formation so that a new phase is stabilized for PtSb$_{1.4}$Sn$_{0.6}$ \cite{Jianjun}. Most importantly, we observe giant anisotropy in $S^2\sigma$ when the heat flow is applied along $a$- or $b$-axis of the crystal structure $S^2\sigma$ is much larger when compared to $c$-axis.

\section{ASSOCIATED CONTENT}

\textbf{Accession Codes}\\
The corresponding crystallographic information file (CIF) has been deposited with the Cambridge Crystallographic Database Centre (CCDC) and can be obtained free of charge via\\ http://www.ccdc.cam.ac.uk/conts/retrieving.html (or from the CCDC, 12 Union Road, Cambridge CB2 1EZ, UK; Fax:  +44-1223-336033; E-mail: deposit@ccdc.cam.ac.uk - depository number 2181562).

\section{AUTHOR INFORMATION}

\textbf{Corresponding Authors}\\
Email:yuliu@lanl.gov\\
Email:petrovic@bnl.gov\\
\textbf{Present Addresses}\\
$^\sharp$ Los Alamos National Laboratory, Los Alamos, NM 87545, USA\\
$^\P$ ALBA Synchrotron Light Source, Cerdanyola del Valles, E-08290 Barcelona, Spain\\
\textbf{ORCID}\\
Yu Liu: 0000-0001-8886-2876\\
Svilen Bobev: 0000-0002-0780-4787\\
Cedomir Petrovic: 0000-0001-6063-1881\\
\textbf{Notes}\\
The authors declare no competing financial interest.

\section{ACKNOWLEDGEMENTS}

Work at BNL is supported by the Office of Basic Energy Sciences, Materials Sciences and Engineering Division, U.S. Department of Energy (DOE) under Contract No. DE-SC0012704. This research used the 8-ID beamline of the NSLS II, a U.S. DOE Office of Science User Facility operated for the DOE Office of Science by BNL under Contract No. DE-SC0012704. The work carried out at the University of Delaware was supported by the U.S. Department of Energy, Office of Science, Basic Energy Sciences, under Award No. DE-SC0008885.\\

Synopsis

\begin{figure}
  \centerline{\includegraphics[scale=1]{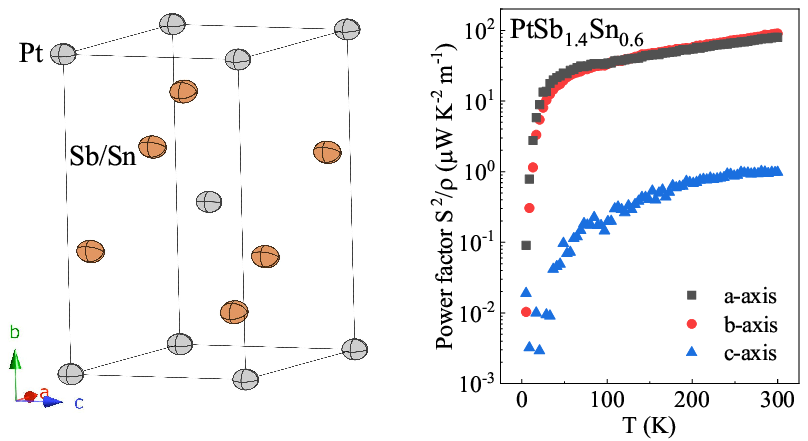}}
\end{figure}

   PtSb$_{1.4}$Sn$_{0.6}$ single crystals crystallize in orthorhombic marcasite structure even though both PtSb$_2$ and PtSn$_2$ are cubic materials at ambient pressure. The alloying creates not only simple random distribution of atoms but also significant change in bond energies and ethalpies of formation so that a new phase is stabilized for PtSb$_{1.4}$Sn$_{0.6}$. Interestingly, we observe giant anisotropy in thermoelectric power factor $S^2\sigma$. When the heat flow is applied along $a$- or $b$-axis of the crystal structure $S^2\sigma$ is much larger when compared to $c$-axis.

\end{document}